\documentclass[aps, prb, reprint, twocolumn, superscriptaddress]{revtex4-2}

\usepackage{graphicx}
\usepackage[cmyk,dvipsnames]{xcolor}
\usepackage{amsbsy}
\usepackage{amsmath}
\usepackage{amssymb}
\usepackage{amsfonts}
\usepackage{bigints}
\usepackage[utf8]{inputenc}
\usepackage[T1]{fontenc}
\usepackage{placeins}
\usepackage{float}
\usepackage{hyperref}
\hypersetup{
    colorlinks,
    linkcolor={blue},
    citecolor={blue},
    urlcolor={blue}
}

\makeatletter
  \def\my@tag@font{\normalsize}
  \def\maketag@@@#1{\hbox{\m@th\normalfont\my@tag@font#1}}
  \let\amsmath@eqref\eqref
  \renewcommand\eqref[1]{{\let\my@tag@font\relax\amsmath@eqref{#1}}}
\makeatother

\renewcommand{\vec}[1]{\mathbf{#1}}
\newcommand{\nvec}[1]{\hat{\vec{#1}}}

\allowdisplaybreaks

\begin{document}

\title{Stability of Hopfions in Bulk Magnets with Competing Exchange Interactions}

\newcommand{\fz}{Peter Gr\"unberg Institut and Institute for Advanced Simulation, Forschungszentrum J\"ulich and JARA, 52425 J\"ulich, Germany}
\newcommand{\iceland}{Science Institute and Faculty of Physical Sciences, University of Iceland, VR-III, 107 Reykjav\'{i}k, Iceland}
\newcommand{\rwth}{Department of Physics, RWTH Aachen University, 52056 Aachen, Germany}
\newcommand{\stpetersburg}{Department of Physics, St. Petersburg State University, 198504 St. Petersburg, Russia}
\newcommand{\itmo}{Russian Faculty of Physics and Engineering, ITMO University, 197101 St. Petersburg, Russia}

\author{Moritz Sallermann}
	\email[]{m.sallermann@fz-juelich.de}
 	\affiliation{\fz}
	\affiliation{\iceland}
 	\affiliation{\rwth}

\author{Hannes J\'{o}nsson}
 	\affiliation{\iceland}

\author{Stefan Blügel}
 	\affiliation{\fz}

\date{\today}

\begin{abstract}
Magnetic hopfions are string-like three-dimensional topological solitons, characterised by the Hopf invariant. They serve as a fundamental prototype for three-dimensional magnetic quasi-particles and are an inspiration for novel device concepts in the field of spintronics. Based on a micromagnetic model and without considering temperature, the existence of such hopfions has been predicted in certain magnets with competing exchange interactions. However, physical realisation of freely moving hopfions in bulk magnets have so far been elusive. Here, we consider an effective Heisenberg model with competing exchange interactions and study the stability of small toroidal hopfions with Hopf number $Q_\text{H}=1$ by finding first-order saddle points on the energy surface representing the transition state for the decay of hopfions via the formation of two coupled Bloch points. We combine the geodesic nudged elastic band method and an adapted implementation of the dimer method to resolve the sharp energy profile of the reaction path near the saddle point. Our analysis reveals that the energy barrier can reach substantial height and is largely determined by the size of the hopfion relative to the lattice constant.
\end{abstract}

\maketitle

\section{Introduction}
During the past decade, the topological classification of the electronic and magnetic structures has been an important subject in condensed matter physics. On the magnetic side, the nanoscale magnetic skyrmion has been at the center of attention. It is a localized magnetization texture $\vec{n}(x,y)$ with particle like properties described as a two-dimensional (2D), topological soliton of finite energy. The topology is characterized by the second homotopy group $\pi_2(\mathbb{S}^2)={Q}$, where the winding number $Q$, determined through the unity magnetization field $\vec{n}(x,y)$ as 
\begin{equation}
    Q[\vec{n}] = \frac{1}{4\pi}\int_{\mathbb{R}^2} \vec{n} \cdot \left(\frac{\partial \vec{n}}{\partial x} \times \frac{\partial \vec{n}}{\partial y}\right) \mathrm{d} \vec{r}\, ,
    \label{eq:skyrmion_charge}
\end{equation}
and also referred to as topological charge, is the additive group of integers. As a consequence of $\pi_2(\mathbb{S}^2)={Q}$, it is possible to split the set of all maps $\vec{n}(x,y): \mathbb{S}^2\longrightarrow \mathbb{S}^2$ from the 2-sphere of the base manifold $\vec{r}_{\scriptscriptstyle\Vert}=(x,y)\in \mathbb{R}^2$, properly embedded in $\mathbb{R}^2\cup \{\infty\} \longleftrightarrow \mathbb{S}^2$ by the natural boundary condition that $\vec{n}(|\vec{r}_{\scriptscriptstyle\Vert}|\rightarrow\infty)=\nvec{e}_z$ goes to the ferromagnetic state at large distances, onto the 2-sphere of the magnetization field described as a unity field $\mathbf{n}(\vec{r}_{\scriptscriptstyle\Vert})\in \mathbb{S}^2$ with $\mathbf{n}\in \mathbb{R}^3$ and $|\mathbf{n}|=1$, into homotopically distinct classes that cannot be continuously deformed into each other.
Usually one identifies a magnetic skyrmion by the topological property that the magnetization pattern is twisted in such a way that it points in all possible directions exactly once resulting in the winding number $Q=-1$. Here, $\nvec{e}_z$ is the unit vector of the magnetization in z-direction. The magnetic skyrmion received its name from its topological similarity to the skyrmion in the original Skyrme model~\cite{Skyrme:62}, where the topological soliton describes a three-dimensional (3D) particle in the pion field and is characterized by the third homotopy group $\pi_3(\mathbb{S}^3)={Q}$ satisfying the mapping $\mathbb{S}^3\longrightarrow \mathbb{S}^3$~\cite{manton_topological_2004}.

The solitonic stability of the magnetic skyrmion goes back to the micromagnetic energy functional for which stable  skyrmions~\cite{bogdanov_thermodynamically_1989,melcher_chiral_2014} were conjectured in chiral magnets stabilized by the presence of the relativistic Dzyaloshinskii-Moriya (DMI)~\cite{moriya_anisotropic_1960} interaction. The two-dimensionality, the nanoscale size, the emergence in thin films ~\cite{Heinze_11,romming:13} and heterostructures~\cite{Moreau-Luchaire2016,soumyanarayanan2017}, the stability~\cite{bessarab_method_2015,vonMalottki:2017,bessarab:2018,varentcova:2018,hoffmann2020}, the possible movement by low electric current densities~\cite{jonitz2010,woo2016observation} and the electrical detectability~\cite{Neubauer:09,Hanneken2015,Crum2015,Maccariello2018} turned the skyrmion into an interesting and intensively studied entity for information storage~\cite{fert_skyrmions_2013}, processing~\cite{Zhang2020} and neuromorphic computing~\cite{Pinna:20,Finocchio:21}. Extensions to antiskyrmions~\cite{Koshibae:16,hoffmann2017,kovalev:18}, skyrmions in centrosymmetric magnets stabilized by exchange frustration~\cite{Leonov2015,Kurumaji:19,Bouaziz:22} or skyrmions with higher topological charges~\cite{Kuchkin:2020} have been discussed.

Less advanced is our understanding of these 2D topological objects in 3D solids, and indeed skyrmions exist also in chiral magnetic bulk crystals~\cite{Muehlbauer:09,yu_real-space_2010,Yu2011}.
On the one hand, we expect here the topologically trivial embedding of the 2D skyrmions as 3D skyrmion tubes or strings with large aspect ratios building complex three-dimensional filamentary magnetic textures in bulk materials. In fact, such rope-like twisted patterns of skyrmion strings were only confirmed recently~\cite{Zheng2021}. On the other hand, also the emergence of 3D localized hybrid magnetization textures consisting of a smooth skyrmion (tube) field terminated by one, as in the case of the chiral bobber~\cite{rybakov_three-dimensional_2013,rybakov_new_2015}, or two monopoles or Bloch points, respectively, as in the case of the chiral magnetic globule~\cite{muller_coupled_2020}  (also coined toron as in Ref.~\cite{li_mutual_2022}) are theoretically conceivable and have lately been confirmed experimentally~\cite{zheng_experimental_2018}. The hybrid nature of these localized particles consisting of a smooth and a singular magnetic texture leads to electric transport properties very different from the smooth skyrmion textures~\cite{Redies:19}.

\begin{figure}
    \centering
    \includegraphics[width=\linewidth]{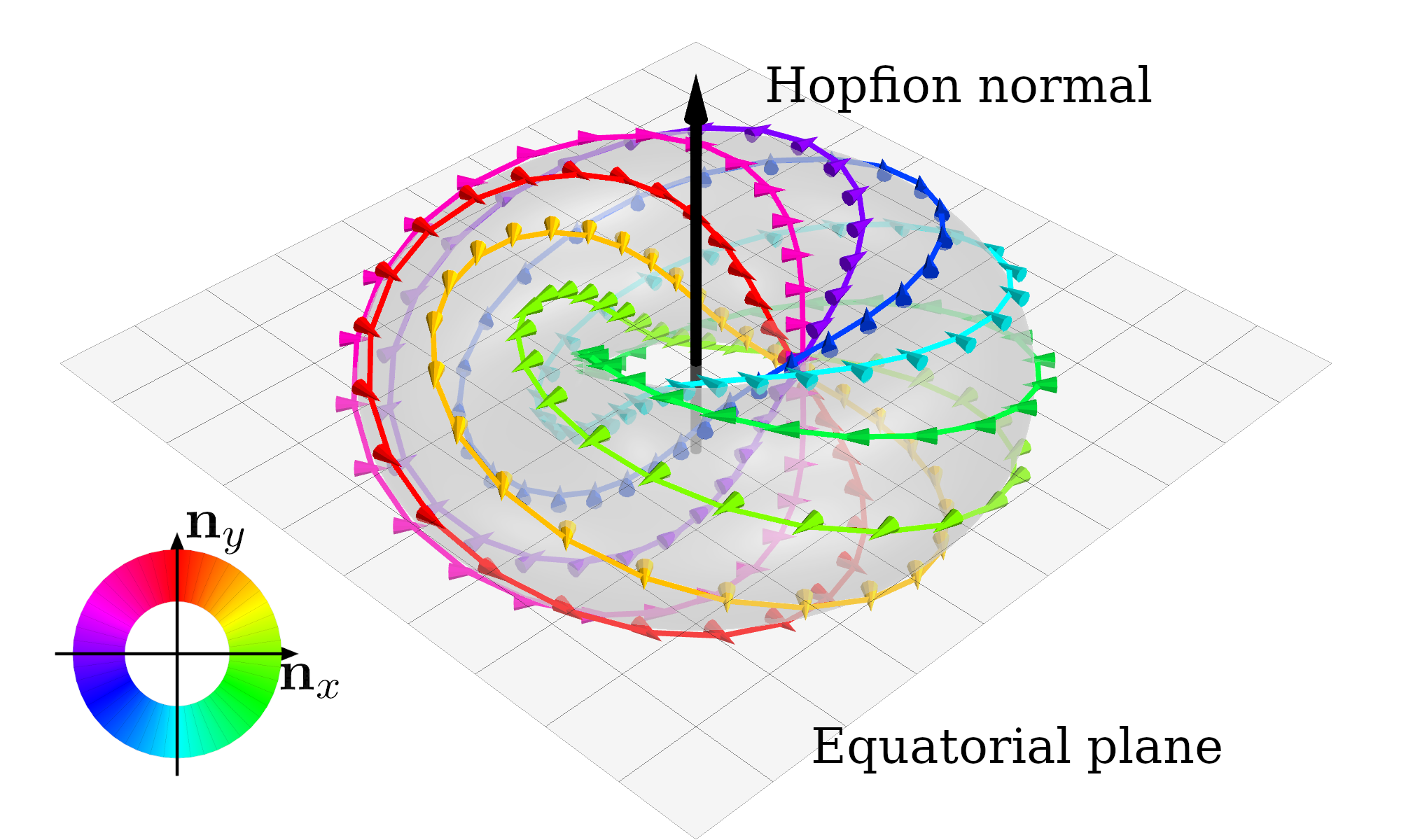}
    \caption{Illustration of the geometry of an isotropic toroidal hopfion with unit Hopf invariant. The toroid is formed by the $n_z=0$ isosurface of the spin direction $\vec{n}(\vec{r})$. Further indicated are the equatorial plane and one of its normals passing through the center point of the toroid. The latter is called the hopfion normal. The colored curves (and cones) are depictions of pre-images, spatial curves along which the spin direction is constant and points into the direction of the cone tip. The color is chosen according to the spin orientation as indicated in the inset. Since this hopfion has a Hopf invariant of $Q_\mathrm{H} = 1$, each pair of pre-images is linked exactly once.}
    \label{fig:hopfion_illustration}
\end{figure}

In this work, we go one step further and focus on topological Hopf solitons or hopfions, smooth localized fully 3D magnetization textures $\vec{n}(\vec{r})$, with $\vec{r}\in\mathbb{R}^3$ and $\vec{n}\in\mathbf{S}^2$ satisfying the first Hopf map~\cite{hopf_uber_1931}  $\vec{n}:\mathbb{S}^3\longrightarrow \mathbb{S}^2$ with the related homotopy group $\pi_3(\mathbf{S}^2)=Q_\mathrm{H}$ employing the analogous compactification principle for the base manifold $\mathbb{R}^3\cup \{\infty\} \longleftrightarrow \mathbb{S}^3$ as for the skyrmions above. Obviously, the map $\vec{n}$ is surjective and for any point $\vec{p}\in \mathbb{S}^2$ on the magnetization 2-sphere its pre-image, $\vec{n}^{-1}(\vec{p})$, is a curve in the base manifold and homeomorphic to a circular fibre $\mathbb{S}^1\in \mathbb{S}^3$ of constant spin direction $\vec{p}$. Two distinct points $\vec{p}$ on $\mathbb{S}^2$ lead to two distinct circles $\mathbb{S}^1$, which are interlinked and the integer, $Q_\mathrm{H}(\vec{n})$, which counts the number of links of two mutual pre-images, is known as the Hopf index $Q_\mathrm{H}(\vec{n})$. It is the defining topological property of the hopfion. While for the skyrmion the magnetization points in every
possible direction exactly once, for the hopfion the magnetization points in any fixed direction on closed spatial curves.
In Fig.~\ref{fig:hopfion_illustration} the example of a toroidal hopfion with Hopf number $Q_\mathrm{H}=1$ is shown.
The Hopf index can be computed directly from the magnetization field using the Whitehead formula ~\cite{whitehead_expression_1947}
\begin{equation}
    Q_\mathrm{H}[\vec{n}] = -\frac{1}{(8\pi)^2} \int_{\mathbb{R}^3} \vec{F} \cdot \vec{A} \;\mathrm{d}\vec{r}\, ,
    \label{eq:hopf_charge}
\end{equation}
where the two auxiliary 3D vector fields, the solenoidal field $\vec{F}$ defined as
\begin{equation*}
    {F}_\alpha = \varepsilon_{\alpha\beta\gamma} \vec{n} \cdot \left( \frac{\partial \vec{n}}{\partial r_\beta} \times \frac{\partial \vec{n}}{\partial r_\gamma}\right)\quad\text{with}\quad\alpha,\beta,\gamma\in\{x,y,z\}\, 
\end{equation*}
and the implicitly defined field $\vec{A}$, $\nabla \times \vec{A} = \vec{F}$, have relations familiar from electrodynamics. Here, $\varepsilon_{\alpha\beta\gamma}$ is the antisymmetric 3D Levi-Civita tensor. While single  skyrmions can propagate freely in 2D, hopfions can propagate in all three spatial directions, which can open up an additional dimension in the development of magnetic storage, data processing systems, or neuromorphic devices.

Stable static nanoscale magnetic hopfions with and without external magnetic field, magnetic anisotropies or device boundary conditions have been theoretically proposed in several types of magnets with suggestions of different hopfion stabilization mechanisms. Most straightforward seems the generation of hopfions from skyrmions in chiral ferromagnets~\cite{sutcliffe2017,sutcliffe_hopfions_2018,Liu:18,Tai:18,li_mutual_2022} or chiral magnets with helical or conical background~\cite{Voinescu:20}. Examples include the creation of a closed loop of a twisted magnetic skyrmion string~\cite{sutcliffe2017} or from a target skyrmion in a nanocylinder~\cite{sutcliffe_hopfions_2018, Liu:18} by
introducing perpendicular magnetic anisotropy at interfaces. Currently the stabilization of hopfions in all-Heisenberg magnets with frustrated exchange interactions proposed by several authors~\cite{bogolubsky:88,sutcliffe_skyrmion_2017,Barts2021,rybakov_magnetic_2019} looks most promising. Furthermore, there is the group of beyond Heisenberg magnets where higher-order spin-interactions lead to the stabilization of hopfions. An example is the Faddeev-type~\cite{Faddeev1997} hopfions proposed in magnets with strong topological orbital magnetism~\cite{Grytsiuk2020}. Common to these models and mechanisms is that the description of the physics, when expressed in terms of micromagnetic energy functionals, leads to classical field equations with higher order spatial derivatives that lower the energy of 3D twisted magnetization textures and stabilize hopfions as local energy minima in the topological sector of finite $Q_\mathrm{H}$.

Despite these theoretical works, and despite the report by Kent \textit{et al.}~\cite{Kent2021} of strong experimental evidence for the formation of a hopfion confined in a nanodisk of magnetic multilayers with strong DMI and out-of-plane magnetic anisotropy, a free single hopfion is still very elusive experimentally. This fact raises questions about the proper materials, the experimental characterization capabilities of nanoscale 3D magnetization textures and the stability of hopfions. Is the topological protection by the topological linking number against the decay to the ferromagnetic ground state weaker than the protection of the skyrmion by the topological charge?

One source of instability for any topologically protected magnetization particle is the granularity of the underlying crystalline lattice. The concept of topological protection exclusively applies to a smooth field $\vec{n}(\vec{r})$, which is a very good approximation to the more realistic description by a spin-lattice model, where the magnetization texture is described by local magnetic moments $\{\vec{n}_i\}$ placed at atomic sites $i$, when the difference of angles between neighboring spins is small compared to their distance. This becomes particularly an issue when the magnetization particle shrinks at the saddle point to the ferromagnetic transition~\cite{hoffmann2020, bessarab:2018, varentcova:2018}. A second source of instability is the formation of (pairs of) Bloch points~\cite{hoffmann2020} in the vicinity of saddle points. In 3D, the phase space volume for the formation of Bloch points is larger than in 2D, and accordingly more pairs can be formed, which may indicate that hopfions may have more decay channels than skyrmions. On the other hand, at least for chiral hopfions in constraint geometries, a sizeable barrier was calculated for the transition to a magnetic toron~\cite{li_mutual_2022} under an external magnetic field, which is consistent with the experimental result of Kent \textit{et al.}~\cite{Kent2021}.   

In this work we investigate the stability of free isolated atomic scale hopfions in ferromagnets with frustrated exchange interactions against the decay to the ferromagnetic ground state. In order to shed light onto this uncharted territory, we focus on the stability of the simplest possible hopfion, a toroidal hopfion with Hopf number $Q_\mathrm{H}=1$. The magnet with exchange frustration -- a situation where the magnetic state does not energetically satisfy the energy minimization of all mutual ferro- or antiferromagnetic pair interactions between local magnetic moments at different sites -- is described by a Heisenberg model with competing exchange interactions.

We explore the variation of the energy barrier between the topologically nontrivial and trivial magnetic state as well as the hopfion nucleation energy, respectively, for a large set of exchange parameters. These energy barriers are calculated by finding first-order saddle points, which are stationary points in a high-dimensional energy landscape with exactly one mode of decay instability, or in other words, of negative energy curvature. They are maxima of minimum energy paths (MEPs) and therefore provide information about the energy bottleneck that the system has to overcome in a given decay trajectory. These energy barriers enter into thermodynamical models of rare events such as transition state theory~\cite{bessarab_harmonic_2012}, which provide an estimate of the thermal stability of hopfions. 

To calculate the energy barriers for the hopfion decay efficiently, we combine the geodesic nudged elastic band (GNEB) method~\cite{bessarab_method_2015}, which has been extensively used in the study of skyrmions~\cite{vonMalottki:2017,bessarab:2018,varentcova:2018,hoffmann2020}, with our implementation of a dimer method~\cite{henkelman_dimer_1999}. Our variant of the dimer method is a modification of the GNEB method. It is a single ended saddle point search method similar to minimum mode following, which has also been applied to skyrmions~\cite{muller_duplication_2018}.

The Heisenberg parameters span a gigantic phase space of stable and excited magnetic structures, of which hopfions may cover very particular regions. In order to target and explore at least one hopfion region successfully with Heisenberg parameters, we make use of a recently developed micromagnetic model by Rybakov \emph{et al.}~\cite{rybakov_magnetic_2019} for cubic magnets with frustrated exchange interactions, negligible magnetic anisotropy and absence of an external magnetic field, which exhibits hopfions as energy minima under particular conditions of the micromagnetic parameters. The micromagnetic parameter space for which hopfions are stable provides the key to our choice of effective Heisenberg exchange parameters. For convenience, this model is referred to as RKBDMB-model throughout the paper. 

To our surprise, we found that the energy barrier for the set of parameters studied can reach a significant fraction of the hopfion energy. In general, we can conclude that, when working in the regime of (small) toroidal hopfions, the energy barrier increases with the hopfion size or, equivalently, when the angle between magnetic moments on neighboring atomic sites decreases.

The article is organised as follows: First, we summarise the RKBDMB-model. Furthermore, we introduce a rescaling of the RKBDMB-model in terms of reduced parameters to simplify our computational analysis. Second, we introduce the Heisenberg model. Details of all  exchange parameters used can be found in Appendix~\ref{App:Exchange_Parameters}. In chapter~\ref{sec:Computational}, we describe the computational method for estimating the stability of the hopfions in terms of the energy barrier between the hopfion state and the ferromagnetic ground state. This analysis is performed by means of atomistic spin simulations employing the Heisenberg model with exchange parameters based on the RKBDMB-model. Details of the hopfion ansatz function we used to begin the energy minimization of the hopfion configuration for the Heisenberg parameters is given in Appendix~\ref{App:Ansatzfunction}. Then, we introduce and discuss two numerical methods, (i) the geodesic nudged elastic band (GNEB) method and (ii) a new formulation of the dimer method to investigate the details of the energy barrier. In the results part, we apply these methods to toroidal hopfions with hopfion number $Q_\mathrm{H}=1$ and discuss the findings. This is followed by the conclusion section. The color code used to represent the vector field direction is defined in Appendix~\ref{App:Colorcode}, and the structure of the magnetic globules, encountered in the hopfion decay, is described in Appendix~\ref{App:Globule_configurations}.

\section{Models}
\subsection{Micromagnetic Model}
We briefly summarize the RKBDMB-model~\cite{rybakov_magnetic_2019}, describing magnetization textures in terms of the unit vector field $\vec{n}(\vec{r})$, with $|\vec{n}(\vec{r})|=1$, in cubic bulk magnets with frustrated exchange and negligible magnetic anisotropy by the energy functional 
\begin{widetext}
\begin{equation}
        E=\bigintss\limits_{\mathbb{R}^3}
        \left[ \mathcal{A} 
        \sum_\alpha
        \Bigg(\! \frac{\partial \vec{n}}{\partial r_\alpha}  \!\Bigg)^{\!\!2}
        \!\!\right. \left. +\,\mathcal{B}\!
        \sum_{\alpha,\beta\neq\alpha} 
        \Bigg(\!\frac{\partial^2 \vec{n}}{\partial r_\alpha^2} - \frac{\partial^2 \vec{n}}{\partial r_\beta^2} \! \Bigg)^{\!\!2}\right. 
        \!\! \left.  +\, \mathcal{C}\! 
        \sum_{\alpha,\beta\neq\alpha}  
        \Bigg(\! \frac{\partial^2 \vec{n}}{\partial r_\alpha \partial r_\beta} \!\Bigg)^{\!\!2}\,\right] 
        \mathrm{d}{\vec{r}}\, ,
    \label{eq:energy_continuous}
\end{equation}
\end{widetext}
where $\alpha, \beta \in \{x,y,z\}$, $\mathcal{A}$ is the spin-stiffness, and  $\mathcal{B}$, $\mathcal{C}$ are exchange constants beyond the micromagnetic standard model.  

To simplify our analysis it is convenient to introduce a system of reduced units. Via the substitutions
\begin{equation}
    \vec{r} \rightarrow \vec{x} = \vec{r}/r_0 \quad\text{with}\quad r_0 = \sqrt{\frac{\mathcal{B} + \mathcal{C}}{\mathcal{A}}}
    \label{eq:reduced_r}
\end{equation}
and
\begin{equation}
    E \rightarrow \mathcal{E} = E/E_0 \quad\text{with}\quad E_0 = \mathcal{A}r_0,
    \label{eq:reduced_e}
\end{equation}
\eqref{eq:energy_continuous} is transformed into
\begin{widetext}
\begin{equation}
         \mathcal{E} = \bigintss_{\mathbb{R}^3}
         \left[\sum_{\alpha} \left(\!\frac{\partial \vec{n}}{\partial x_\alpha} \!\right)^2 \right. \left. +\, (1-\gamma)\! \sum_{\alpha,\beta\neq\alpha} \left(\! \frac{\partial^2 \vec{n}}{\partial x_\alpha^2} - \frac{\partial^2 \vec{n}}{\partial x_\beta^2} \!\right)^2\right.
         \left. + \, \gamma\! \sum_{\alpha,\beta\neq\alpha} \left(\! \frac{\partial^2 \vec{n}}{\partial x_\alpha \partial x_\beta} \!\right)^2\right] \mathrm{d}{\vec{r}},
    \label{eq:energy_continuous_reduced}
\end{equation}
\end{widetext}
with one dimensionless parameter
\begin{equation}
    \gamma = \frac{\mathcal{C}}{\mathcal{B}+\mathcal{C}}\,.
    \label{eq:reduced_gamma}
\end{equation}
We restrict the analysis to the regime of $\mathcal{A}, \mathcal{B}, \mathcal{C} > 0$ in which the ground state is ferromagnetic and consequently $\gamma$ only takes values within the interval $[0,1]$. It is evident that $\gamma = 0$ corresponds to $\mathcal{C} = 0$, while $\gamma = 1$ corresponds to $\mathcal{B} = 0$. Further below, we show that $\gamma$ can be interpreted as an anisotropy parameter that controls the symmetry of locally optimal spin configurations, ranging from (amongst others) toroidal hopfions with hexagonal symmetry at $\gamma=0$ to the fully isotropic case attained at $\gamma = 6/7$ when $\mathcal{C} = 6\mathcal{B}$, to a quadratic symmetry at $\gamma=1$. The parameter $r_0$ is a length scale that isotropically scales the size of these configurations without changing the symmetry. 

According to the approximate criterion, derived by Rybakov \textit{et al.}~\cite{rybakov_magnetic_2019}, hopfions are likely to be local energy minima of discrete realisations of the energy functional~\eqref{eq:energy_continuous_reduced}, if the condition
\begin{equation}
    \mathrm{max}\big(\gamma, 6(1-\gamma)\big) \gtrapprox 6.5 \left(\frac{d}{r_0}\right)^2
    \label{eq:stability_criterion}
\end{equation}
holds, where $d$ is the distance between nearest neighbor spins. In the case of the simple cubic lattice, considered in this article, $d$ equals the lattice constant $a$. %

\subsection{Heisenberg Model}
The RKBDMB-model can be understood as the long-wavelength limit of a minimal effective isotropic Heisenberg Hamiltonian 
\begin{equation}
    \mathcal{H} = -\sum_{s=1}^{4} J_s \sum_{\langle i,j \rangle_s} \vec{n}_{i} \cdot \vec{n}_{j}\, ,
    \label{eq:heisenberg_hamiltonian}
\end{equation}
a spin-lattice model considering the magnetic interaction between atoms up to four atomic shells $s$ of nearest neighbor atoms.
The inner sum runs over unique pairs of atoms $\langle i,j \rangle_s$ with site indices $i, j$ for which the classical spin vectors $\vec{n}_i$ and $\vec{n}_j\in \mathbb{R}^3$ of unit length $|\vec{n}_i|=1\,\forall i$, lie in each others $s$-th nearest neighbor shell. Restricting the long-wavelength expansion of the Heisenberg model to the fourth power in the angular change of the magnetic moments between neighbors relates the micromagnetic and the Heisenberg models by a linear mapping $\mathcal{A} \{\mathcal{B},\mathcal{C}\}=\sum_s a_s \{b_s, c_s\}J_s$ between the exchange coupling strengths $J_{s}$ of the four shells of nearest-neighbor atoms and the micromagnetic parameters $\mathcal{A}$ to $\mathcal{C}$ (for details see $\eqref{eq:abc_from_j})$. Obviously, this linear relationship expresses an enormous compression of information and degrees of freedom when treating a micromagnetic model instead of an atomistic one. It is also clear that the relationship between the parameters $\mathcal{A}$ to $\mathcal{C}$ also depends on the underlying atomic lattice.

\section{Atomistic Spin-Simulations: Computational Details and Methods}
\label{sec:Computational}
Our atomistic spin simulations utilize the classical Heisenberg Hamiltonian~\eqref{eq:heisenberg_hamiltonian} and were performed with the \texttt{Spirit} code~\cite{muller_spirit_2019}. It is obvious that the Heisenberg model describes a much richer spectrum of spin-textures than the micromagnetic model, but the search for hopfions requires an excessive amount of calculations in order to explore the high-dimensional parameter space. This is time consuming and at present very difficult.

Instead, to systematically choose exchange parameters $J_s$ of the Heisenberg Hamiltonian for which we can expect stable hopfions, we make use of their existence within the RKBDMB-model by relating the $J_s$ to the set of reduced units given by \eqref{eq:reduced_r}, \eqref{eq:reduced_e} and \eqref{eq:reduced_gamma} by making three free but reasonable choices. (i) We kept the characteristic energy $E_0$ fixed at $1\,\mathrm{meV}$, (ii) set $J_1 = 40\,\mathrm{meV}$ and (iii) chose a simple cubic lattice with a lattice constant $a=1\,\textup{\AA}$.
As long as the truncation error due to  neglecting higher orders of derivatives of the magnetization structure, when converting between the discrete model \eqref{eq:heisenberg_hamiltonian} and the continuous micromagnetic functional \eqref{eq:energy_continuous_reduced}, remains relatively small these choices do not significantly affect the results. After these quantities were fixed, we arrived at a one-to-one mapping from the symmetry parameter $\gamma$ and the characteristic length scale $r_0$ to the exchange couplings $J_s$.
Then, we sampled $\gamma$ and $r_0$ on a regular grid and used the mapping to the exchange couplings to construct the corresponding Heisenberg Hamiltonian.
To obtain toroidal hopfions, with Hopf charge $Q_\mathrm{H} = 1$, the energy of a suitable ansatz (see Appendix~\ref{App:Ansatzfunction}) was minimized up to machine precision with a solver specialised in norm-conserving optimisation ~\cite{ivanov_efficient_2020}. The simulation box contained $64^3$ spins and the boundary conditions were periodic. This box is sufficient to obtain saddle points for isolated toroidal hopfions with a diameter of up to roughly $30\,a$. While some finite size effects, like a numerical anisotropy due to the shape of the simulation box (and the choice of bravais lattice) as well as slight shape deformations in the case of the largest studied Hopfions are present, they do not influence the energy barriers significantly.
An overview of the tested parameters, as well as how they fit to the predictions of the stability criterion \eqref{eq:stability_criterion}, is presented in Fig.~\ref{fig:plot_parameter_grid}(a). Further details can be found in Appendix~\ref{App:Exchange_Parameters}.

The saddle points have been obtained in a two step process. First, the geodesic nudged elastic band (GNEB) method~\cite{bessarab_method_2015}, as implemented in \texttt{Spirit}, has been used as a pre-convergence step until the effective forces on every spin of every image reached values in the order of $10^{-3}\,\mathrm{meV}$. Please note that the definition of the effective forces in the GNEB and the dimer method is given in the next section.
Then, the two consecutive images on the path with the lowest (most negative) energy curvature were determined by forming finite differences. We then applied the dimer method, a modification of the GNEB method that we implemented in the \texttt{Spirit} code for this purpose, to these two points in order to converge the saddle points down to effective forces on the order of $10^{-12}\,\mathrm{meV}$. For the initial path of the GNEB calculation, we used 20 discrete images and interpolated homogeneously from the hopfion to the ferromagnetic state.

In the following, we briefly describe the GNEB and the dimer method. The latter is a modification of the GNEB method and similar, but not identical, to the dimer method by Henkelman \emph{et al.} ~\cite{henkelman_dimer_1999}.

\begin{figure*}[t]
    \centering
    \includegraphics[width=\linewidth]{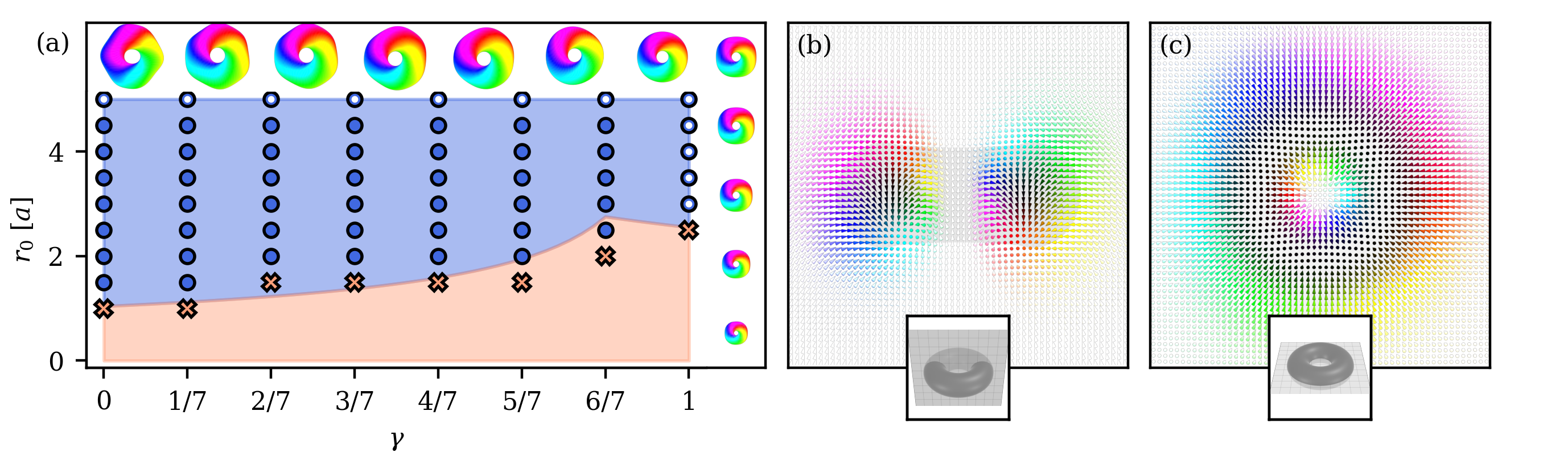}
    \caption{(a) The investigated parameter space ($\gamma$, $r_0$). The characteristic energy $E_0$ has been kept constant at $1\,\mathrm{meV}$, while $\gamma$ has been varied in steps of $1/7$ and $r_0$ in steps of $0.5\,a$, where $a$ is the lattice constant of the simple cubic atomic lattice. The markers show the discrete pairs of ($\gamma$, $r_0$) for which the energy barrier was investigated. A blue circle signifies a stabilized hopfion and a red cross signifies a point where a hopfion could not be stabilized and the system would decay into the ferromagnetic ground state. The blue (red) shaded background illustrates the region where the criterion Eq.~\eqref{eq:stability_criterion}, which is derived from micromagnetic models, predicts hopfion stability (instability). The images of the magnetization textures at the top and right side show $n_z = 0$ isosurfaces of the magnetization texture, for different values of $\gamma$ and $r_0$. The color tone of the surface depends on the azimuthal angle of the spin direction, additionally a detailed depiction of the colorcode can be found in Appendix~\ref{App:Colorcode}. Small white dots on top of the blue circles mark ($\gamma$, $r_0$)-points for which isosurfaces are shown. 
    (b) Cross section of the isotropic ($\gamma$, $r_0$) = (6/7, $5\,a$) hopfion in a plane containing the central normal of the toroid. The bottom inset illustrates the cross section plane in relation to the $n_z = 0$ isosurface. (c) Cross section of the same hopfion in the equatorial plane of the toroid. Again, the bottom inset illustrates the cross section plane in relation to the $n_z = 0$ isosurface.
    }
    \label{fig:plot_parameter_grid}
\end{figure*}

\subsection{Geodesic Nudged Elastic Band Method}
The GNEB method is a scheme for finding minimum energy paths (MEPs) between minima corresponding to initial and final states. To achieve this, the method arranges several images of the spin system in a chain, thereby providing a discretised interpolation between the initial and final configuration. The MEP is found by iteratively following effective forces acting on the entirety of the  chain of images. For a chain of $C$ images with $N$ spins per image, the total force acting on image $\nu$ of the chain is a vector with $3N$ components and given by
\begin{equation}
    F_{\nu} = S_{\nu} + G^{\perp}_\nu \quad\text{with}\quad \nu \in 2,...,C-1,
    \label{eq:force_gneb}
\end{equation}
where $S_\nu$ is called the spring force and $G^{\perp}_{\nu}$ is the gradient force projected orthogonally to the path tangent $t_\nu$. Notice that the endpoints of the chain ($\nu = 1$ and $\nu = C$) are excluded in \eqref{eq:force_gneb} and remain fixed.
The spring force $S_\nu$ ensures that, during the iterative optimization, the discrete images remain distributed equidistantly along the chain and is defined as
\begin{equation}
    S_\nu = \Delta_{\nu,\nu+1} t_{\nu},
\end{equation}
where $\Delta_{\nu,\nu+1}$ is a measure of distance between image $\nu$ and image $\nu+1$ and $t_{\nu}$ is the tangent to the path at image $\nu$. 
The distance between two adjacent images $\nu$ and $\nu+1$ is computed from the angles between spins,
\begin{equation}
    \Delta_{\nu,\nu+1} = \sqrt{\sum\nolimits_i \angle\! \left(\vec{n}^{(\nu+1)}_i , \vec{n}^{(\nu)}_{i}\right)^2 },
\end{equation}
where $\vec{n}^{(\nu)}_i$ is the $i$-th spin of image $\nu$.
The tangent $t_{\nu}$ is found from the central finite difference between image ${\nu+1}$ and image ${\nu-1}$, which is then orthogonally projected to image $\nu$,
\begin{equation}
    t_\nu = 
    \begin{pmatrix}
        \left[\vec{d}^{(\nu-1,\nu+1)}_1 - \left(\vec{d}^{(\nu-1,\nu+1)}_1 \cdot \vec{n}^{(\nu)}_1 \right) \vec{n}^{(\nu)}_1\right]_\alpha\\
        \vdots\\
        \left[\vec{d}^{(\nu-1,\nu+1)}_N - \left(\vec{d}^{(\nu-1,\nu+1)}_N \cdot \vec{n}^{(\nu)}_N \right) \vec{n}^{(\nu)}_N\right]_\alpha\\
    \end{pmatrix},
    \label{eq:path_tangent}
\end{equation}
where
\begin{equation*}
    \vec{d}^{(\nu-1,\nu+1)}_i = \vec{n}^{(\nu+1)}_i - \vec{n}^{(\nu-1)}_i,
\end{equation*}
and $\alpha \in \{x,y,z\}$.
Note that the tangent found in Eq.~\eqref{eq:path_tangent} is additionally normalized as a $3N$ vector.

The gradient force acting on a spin $\vec{n}^{(\nu)}_i$ is found by taking the negative of the partial derivative of the energy, given by the Heisenberg Hamiltonian Eq.~\eqref{eq:heisenberg_hamiltonian}, with regards to the spin direction. The gradient force vector $G_\nu$ is the composition of these single spin forces
\begin{equation}
    G_\nu = 
    \begin{pmatrix}
        - \left[ \nabla_{\vec{n}^{(\nu)}_1} \mathcal{H} \right]_\alpha\\
        \vdots\\
        - \left[ \nabla_{\vec{n}^{(\nu)}_N} \mathcal{H} \right]_\alpha\\ 
    \end{pmatrix},
\end{equation}
and we can form the projected gradient force $G^{\perp}_\nu$ by removing the component parallel to the path tangent
\begin{equation}
    G^{\perp}_\nu = G_\nu - (G_\nu \cdot t_\nu) t_\nu.
\end{equation}

We note that the climbing image variant of the GNEB method (CI-GNEB) can be used to place one of the images exactly at the maximum of the MEP and therefore at a saddle point. In the CI-GNEB method the spring force $S_\nu$, acting on a selected so-called climbing image, is deactivated and replaced by the inverted component of the gradient force along the tangent.

\subsection{Dimer Method}
\begin{figure*}[t]
    \centering
    \includegraphics[width=\linewidth]{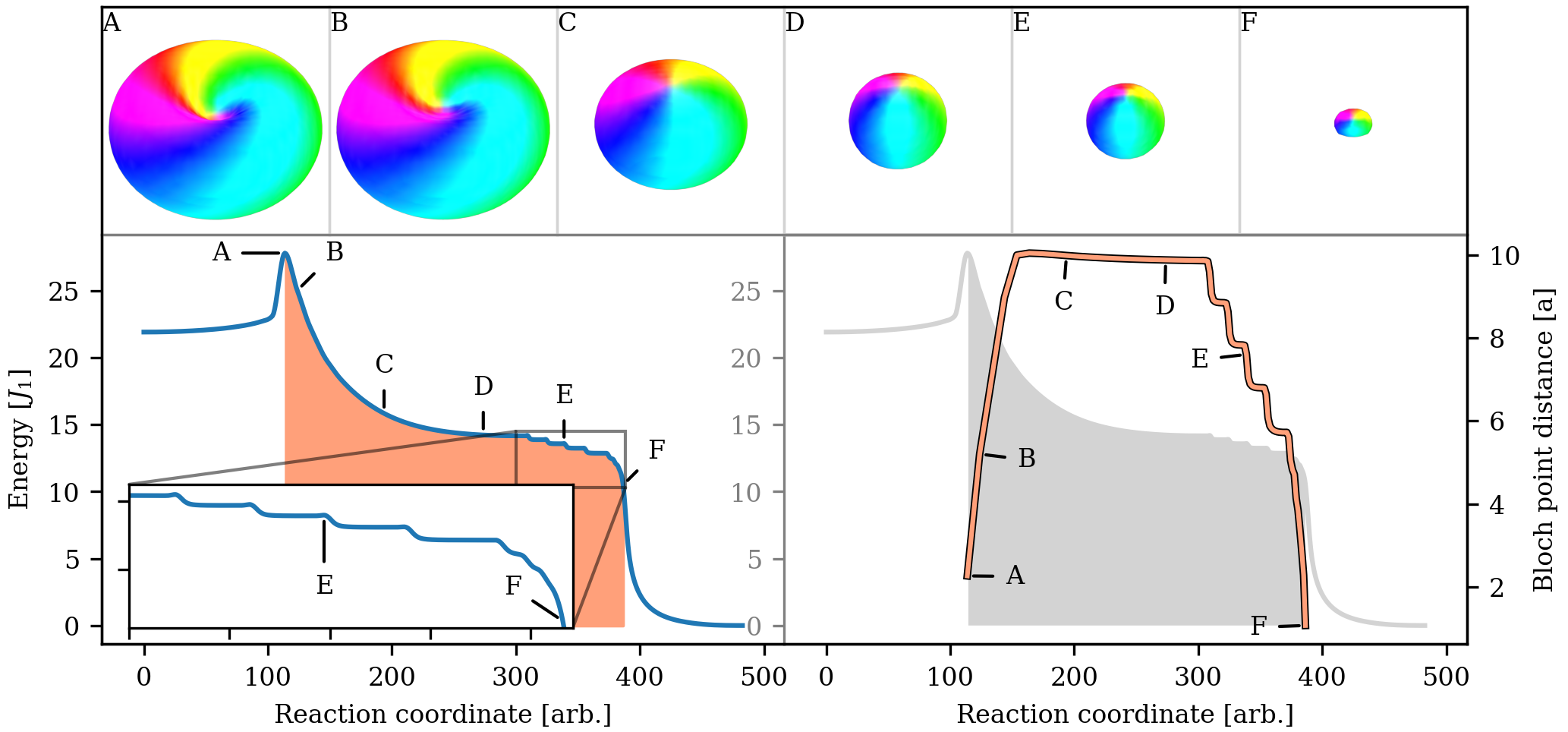}
    \caption{(left) Minimum energy path for $(\gamma, r_0) = (6/7, 5\,a)$. The inset shows a zoomed subsection of the path, where many intermediate shallow minima form. Beyond the saddle point A a pair of Bloch points emerges. The orange shading marks the interval in which this pair exists. (top) The images show renderings of the $n_z = 0$ isosurface at different points of the path -- marked by capital letters. (right) The distance between the pair of Bloch points as a function of the reaction coordinate. The gray shade in the background represents the energy and is intended as a visual guide.}
    \label{fig:meps}
\end{figure*}

One disadvantage of the GNEB method is that both states, the initial and the final, need to be known in advance. Another drawback is that typically the entire path has to be discretized, which makes the calculations  expensive and possibly leads to a low resolution around the saddle point. In this work, the latter made the computations with the GNEB method intractable because the energy barriers for many of the transitions are sharply peaked -- as shown in the example of the path in Fig.~\ref{fig:meps}. One approach to deal with this reduced resolution around the saddle point is to introduce energy weighted spring forces~\cite{henkelman_climbing_2000, ivanov_fast_2021}, which increase the density of images near energy peaks of the path. 

Here, we have chosen an alternative strategy that alleviates both of these problems at once by implementing a variant of the GNEB method that uses only two images, turning the chain into a dimer. 
There are several advantages to this method: On one hand, it allows for arbitrary resolution around the saddle point, and on the other hand the computational effort is greatly reduced since fewer images are needed.
An additional property of the method that deserves mentioning, although we do not make use of it in the present work, is that it (unlike GNEB) does not rely on knowledge of a final state and can, therefore, uncover new transitions and saddle points.

In our dimer method the forces on the two images, $1$ and $2$,  are given by
\begin{equation}
    F_{1/2} = S'_{1/2} + G^{\perp}_{1/2} + T_{1/2}.
\end{equation}
The different force contributions are discussed below.

Compared to the regular GNEB method, the spring forces $S'_{1/2}$ are modified and an equilibrium distance $\delta$ is introduced, which prevents the two images from collapsing into each other, leading to
\begin{equation}
    S'_{1/2} = \pm (\Delta_{1,2} - \delta) t_{1/2}.
\end{equation}
To estimate the path tangents $t_{1/2}$ a forward/backward difference is used, as opposed to the central finite differences from the GNEB method. The definitions of the projected gradient forces $G^{\perp}_{1/2}$ are identical to the regular GNEB method.
Lastly, we define the translational forces $T_{1/2}$ which move the dimer towards an energy maximum along the tangent, without changing the relative distance of the endpoints. They are constructed by inverting the average of the gradient forces of both endpoints and projecting it along the tangent. Before taking the average, the forces are rotated into the tangent frame of the respective spin configuration giving
\begin{align}
    T_{1/2} &= \left[ -\frac{1}{2} \left( G_{1/2} + \mathcal{R}_{2/1 \rightarrow 1/2} G_{2/1}\right)\cdot t_{1/2} \right] t_{1/2},
\end{align}
where $\mathcal{R}_{1/2 \rightarrow 2/1}$ denotes the transformation from the tangent frame of one endpoint configuration into the other.
The transformation $\mathcal{R}_{1 \rightarrow 2} G_{1}$ is found by applying the $3 \times 3$ rotation matrices $R^{(1,2)}_i$, which rotate the $i$-th spin of the first image $\vec{n}^{(1)}_i$ into the $i$-th spin of the second image $\vec{n}^{(2)}_i$, to the component vectors of the gradient force resulting in
\begin{equation}
    \mathcal{R}_{1 \rightarrow 2}  G_1  :=
    \begin{pmatrix}
        - \left[ R^{(1,2)}_1 \nabla_{\vec{n}^{(1)}_1} \mathcal{H} \right]_\alpha\\
        \vdots\\
        - \left[ R^{(1,2)}_N \nabla_{\vec{n}^{(1)}_N} \mathcal{H} \right]_\alpha\\ 
    \end{pmatrix}.
\end{equation}
The rotations $R^{(1,2)}_i$ can be defined by using axes $\vec{n}^{(1)}_i \times \vec{n}^{(2)}_i$ and angles $\angle\left(\vec{n}^{(1)}_i, \vec{n}^{(2)}_i\right)$ together with Rodrigues' formula~\cite{rodrigues_lois_1840}. For the inverse transformation $\mathcal{R}_{2 \rightarrow 1} G_2$ these rotation matrices are transposed.

\begin{figure*}[t]
    \centering
    \includegraphics[width=\linewidth]{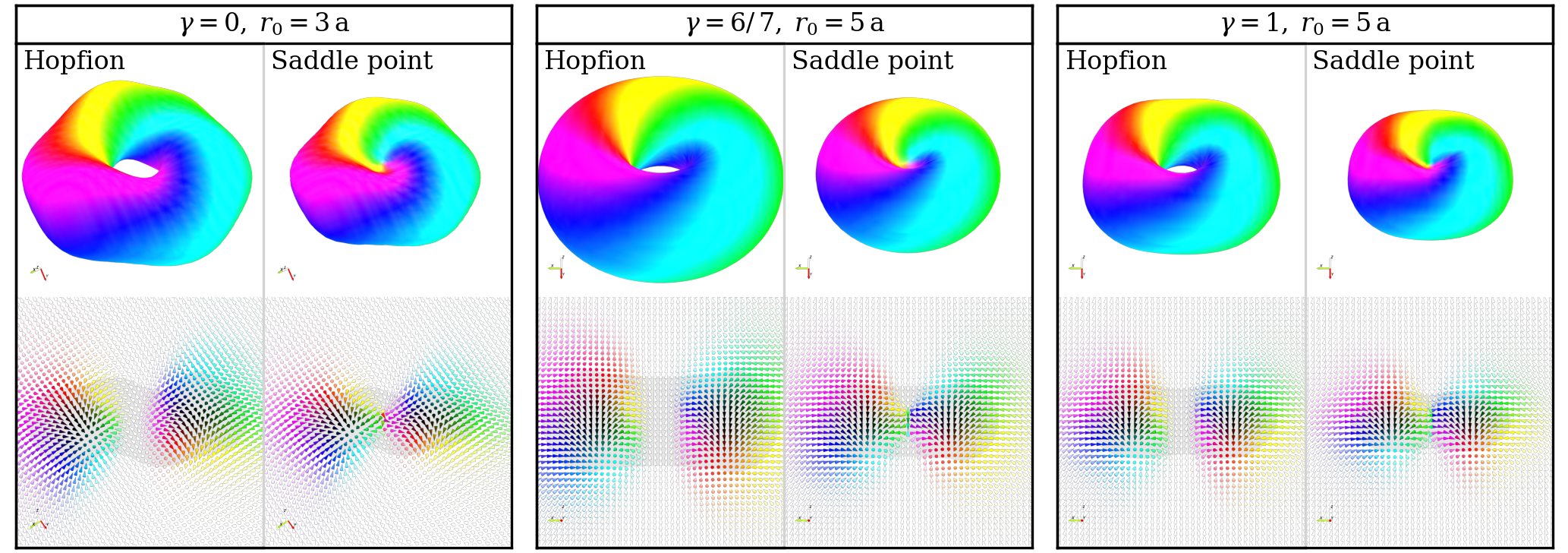}
    \caption{Comparison of several hopfions to their saddle points. The top row contains renderings of the $n_z = 0$ isosurface, while the bottom row shows cross sections in a plane orthogonal to the equatorial plane of the hopfion. The grey shape in the background of the cross sections is a projection of the $n_z = 0$ isosurface and intended as a visual guide. All of the saddle points are a result of the formation of two Bloch points along the hopfion normal.}
    \label{fig:example_saddlepoint}
\end{figure*}

\section{Results and Discussions}
\label{sec:Results}

\begin{figure}[t]
    \centering
    \includegraphics[width=\linewidth]{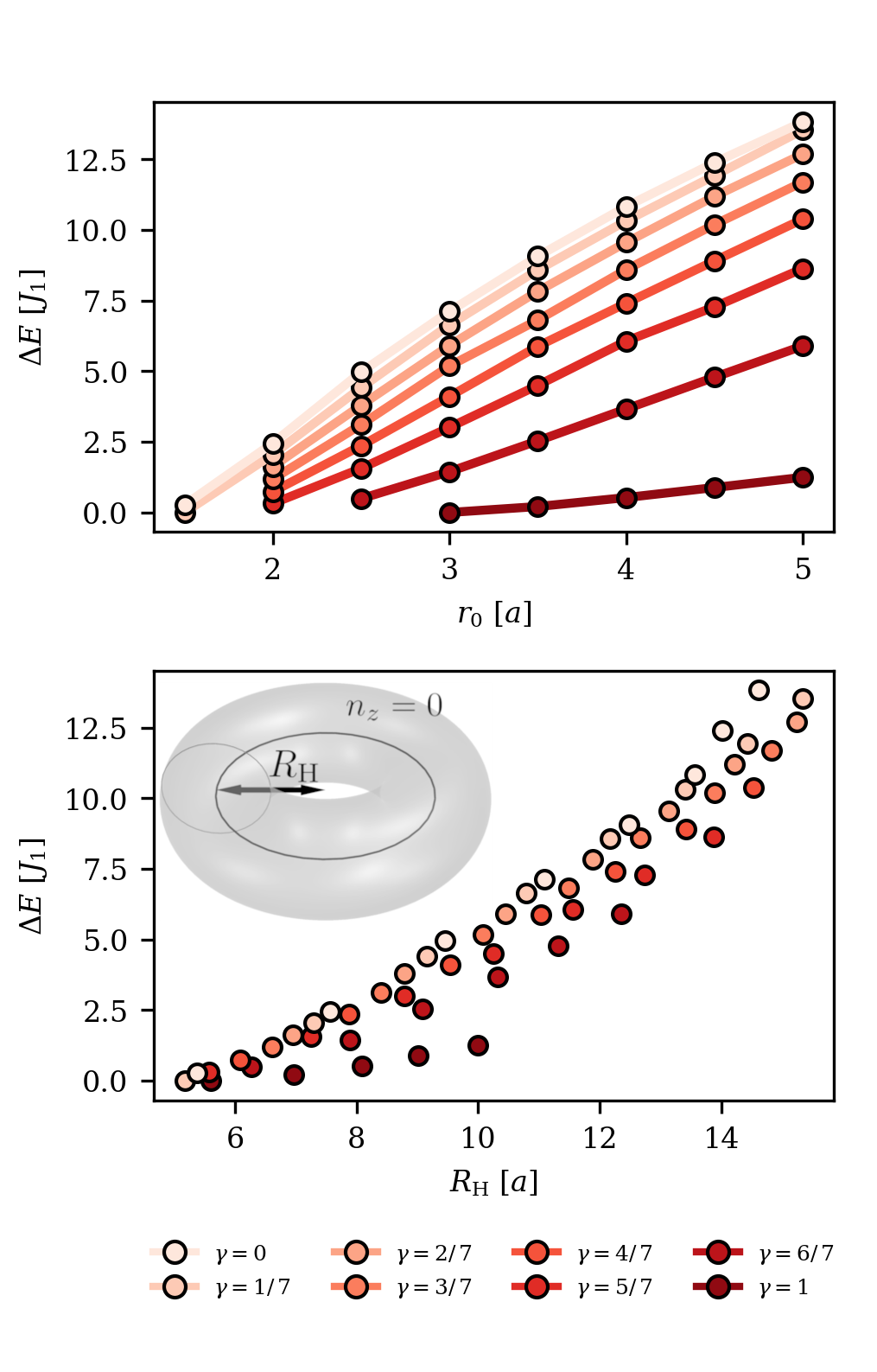}
    \caption{Energy barriers as functions of the characteristic length scale $r_0$ (top panel) and as a function of $R_\mathrm{H}$ (bottom panel). The radius $R_\mathrm{H}$ is related to the overall size of the Hopfion and defined as the average distance of points in the $n_z=0$ isosurface from the toroid center point parallel-projected to the equatorial plane, (bottom panel). The inset in the bottom panel illustrates the definition of $R_\mathrm{H}$.}
    \label{fig:energy_barriers}
\end{figure}

\begin{figure}[t]
    \centering
    \includegraphics[width=\linewidth]{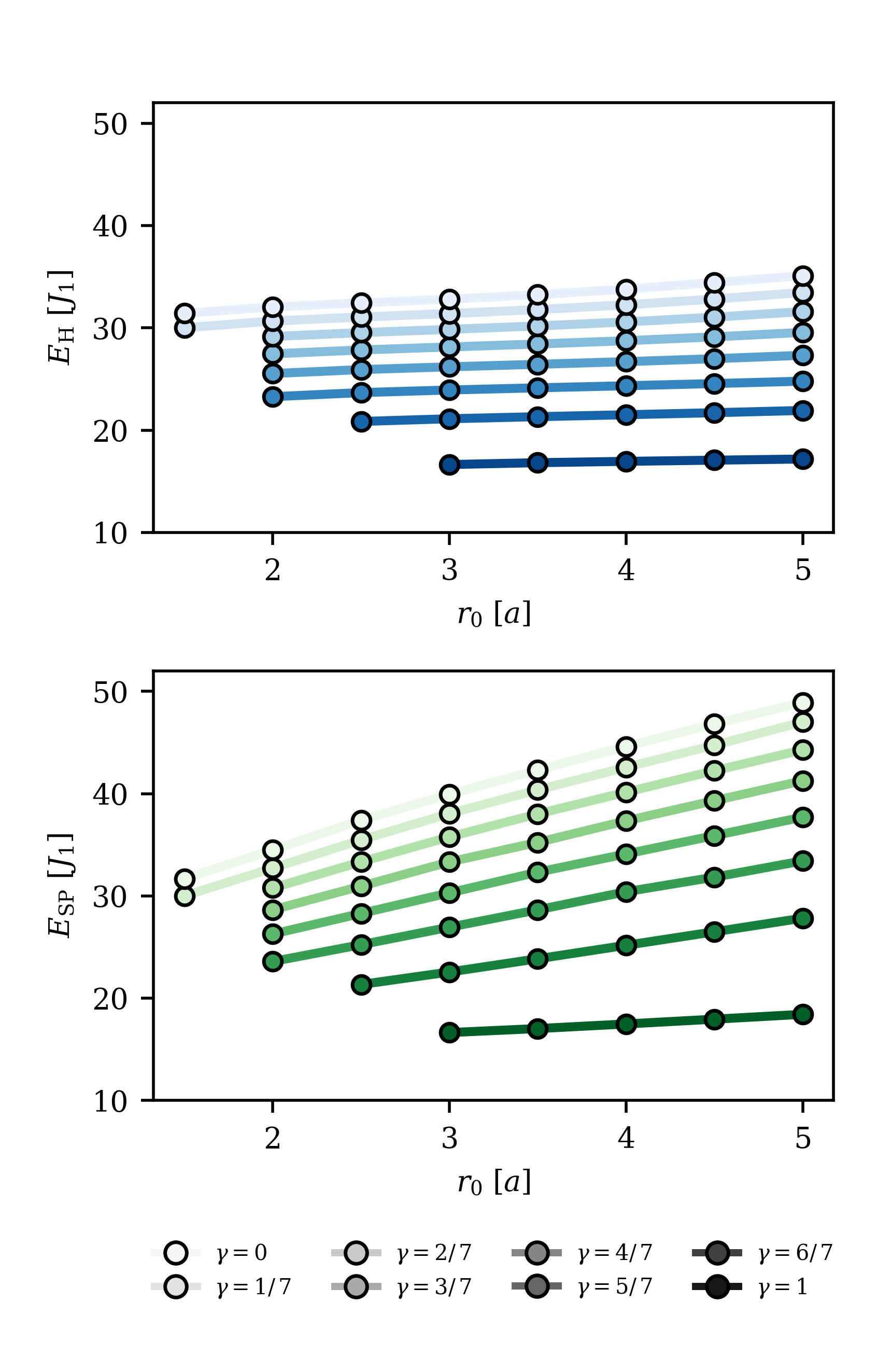}
    \caption{The top (bottom) panel shows the energy difference between the hopfion (saddle point) and the ferromagnetic ground state. Notice the relation between the shade of the points and the value of the symmetry parameter $\gamma$.}
    \label{fig:energetics}
\end{figure}

The computational results of our analysis of the hopfion stability (in the sense of local minima of the Heisenberg Hamiltonian \eqref{eq:heisenberg_hamiltonian}) are summarized in Fig.~\ref{fig:plot_parameter_grid}. Along the perimeter of Fig.~\ref{fig:plot_parameter_grid}(a), the $n_z = 0$ isosurfaces of the magnetization texture are presented which give evidence on the influence of $\gamma$ and $r_0$ on the shape of the stabilized hopfions. A change in the symmetry parameter $\gamma$ deforms the shape of the hopfion taking it from an approximate sixfold symmetry at $\gamma = 0$ to an approximate fourfold symmetry at $\gamma = 1$, while simultaneously reducing the size of the isosurface. For the intermediate value of $\gamma = 6/7$, where the micromagnetic energy functional \eqref{eq:energy_continuous_reduced} is isotropic, the shape of the isosurface is a torus. A change in the characteristic length scale $r_0$, on the other hand, leads to an isotropic growth or shrinkage of the hopfion, preserving its symmetry. Notice that both, $r_0$ and $\gamma$, influence the size of the hopfion.

As Fig.~\ref{fig:plot_parameter_grid}(a) also demonstrates: The analytical stability criterion \eqref{eq:stability_criterion} holds quite well and gives the correct qualitative behavior of larger $r_0$ being needed to stabilize hopfions with fourfold ($\gamma = 1$) rather than sixfold symmetry ($\gamma = 0$). Some small deviations between the criterion and our results can be found at the phase boundary of stability, where the hopfions become too small to sustain. Of course, this boundary is also the regime where the continuous model becomes least accurate.
Fig.~\ref{fig:plot_parameter_grid}(b) and (c) display the directions of individual spins along two different cross-sectional planes of the isotropic $(\gamma, r_0) = (6/7, 5\,a)$ hopfion. Notice how the spin direction transitions to the ferromagnetic state away from the localised hopfion structure. In the cross section orthogonal to the equatorial plane (Fig.~\ref{fig:plot_parameter_grid}(b)) the magnetization winds around the torus with two opposite senses of rotation.

All the saddle points found show the same structure: The ``donut'' hole in the $n_z=0$ isosurface of the hopfion vanishes and the toroid is turned into a biconcave disc (an oblate discus squeezed in the center to obtain two concavities), with pairs of Bloch points emerging along the normal of the equatorial plane of the toroid (hereafter called hopfion normal, see also Fig.~\ref{fig:hopfion_illustration}). A few examples of these saddle points are displayed in Fig.~\ref{fig:example_saddlepoint}.

Knowledge of the saddle point spin configuration is sufficient for the calculation of the energy barriers. It is nonetheless instructive to inspect the minimum energy paths. In Fig.~\ref{fig:meps}, a minimum energy path is displayed for the isotropic hopfion ($\gamma = 6/7$) with the largest tested value of $r_0 = 5\,a$. One noticeable feature of the MEP is the sharp peak of the energy curve at the first saddle point and how quickly the slope rises when moving towards the peak from the initial state. Further, the behavior of the Bloch point pair beyond the saddle point is illustrated in the right panel of Fig.~\ref{fig:meps}, where the distance between the two Bloch points is plotted as a function of the reaction coordinate. After forming at the saddle point, they move apart while the spin structure relaxes into an oblate spheroid. Then, the distance between the pair stays almost constant, while the oblate spheroid becomes more and more spherical, resulting in a local energy minimum -- a globule state. We note that similar states have been reported in chiral magnets by M\"uller \emph{et al.}~\cite{muller_coupled_2020}, although there they do not occur as isolated structures but, instead, are coupled to a spin spiral.

After their creation at the initial saddle point, the Bloch points move towards each other and form many intermittent globule minima, with almost vanishing energy barriers. An enlarged view of that region of the MEP is displayed in the inset of the left panel of Fig.~\ref{fig:meps}. The Bloch point distance between subsequent globule minima changes in steps of one lattice constant, causing the MEP to exhibit a staircase pattern. This, together with the extremely small energy barriers, underlines the strong influence of the crystalline lattice on the the Bloch points forming the globule minima. Finally, the Bloch points annihilate, the remnants of any non-collinear magnetization disappear and the entire object relaxes into the ferromagnetic state.

We note that similar globule configurations do emerge along the MEP for all values of $\gamma$ and $r_0$. However, for the smaller values of $r_0$ tested in this work, these configurations are not necessarily local minima and instead can be observed as shoulders in the MEP, similar to the state labelled 'F' in Fig.~\ref{fig:meps}. In these cases, the initial saddle point directly connects the hopfion to the ferromagnetic ground state without any intermediate minima.

The energy barriers for all the saddle points found are shown in Fig.~\ref{fig:energy_barriers}. In the tested parameter regime, they increase nearly linearly with $r_0$ and reach substantial values for large characteristic length scales $r_0$. When $r_0$ is held constant, hopfions with approximate sixfold symmetry ($\gamma = 0$) show a higher energy barrier than those with approximate fourfold symmetry ($\gamma = 1$). The highest energy barrier found of $13.875\,J_1$ is obtained at $(\gamma,r_0) = (0, 5\,a)$. The height of this energy barrier is about $40\%$ of the energy difference between the hopfion and the ground state.
As already stated, an increase in $\gamma$ decreases the energy barrier. But since the change in $\gamma$ affects the symmetry \textit{and} the size of the hopfion, it is not directly clear which of these effects is the most significant. In order to analyze the role of the hopfion sizes, the bottom panel of Fig.~\ref{fig:energy_barriers} presents the energy barrier against the average radius, $R_\mathrm{H}$, of the $n_z=0$ isosurface measured from the toroid center. If the hopfion is isotropic ($\gamma = 6/7$), $R_\mathrm{H}$ corresponds to the radius of the torus (as also depicted in the inset of the bottom panel of Fig.~\ref{fig:energy_barriers}). A comparison between the two graphs in Fig.~\ref{fig:energy_barriers} reveals that much of the variance between the different $\gamma$ values vanishes and, therefore, hints at the fact that the change in size, and not in symmetry, is the dominant effect.
Additionally, Fig.~\ref{fig:energetics} presents the total energies of the stabilized hopfions as well as the saddle points, both computed as the difference to the energy of the ferromagnetic ground state. While the energy of the hopfion increases only weakly with increasing $r_0$, the energy of the saddle point rises with a steeper slope thus causing the energy barrier, and therefore the stability of the hopfion, to rise together with $r_0$.

Evaluation of the energy barrier for the collapse is the most important aspect of an assessment of the thermal stability of hopfions, but it is also important to estimate the pre-exponential factor in the Arrhenius expression for the rate since it can, in general, vary by several orders of magnitude.
In a concurrent study, Lobanov and Uzdin~\cite{LobanovPreprint} have calculated the pre-exponential factor for the collapse of hopfions, using a similar Hamiltonian to describe the spin interaction, and obtained an estimate of 10$^{18}$ to 10$^{20}$ s$^{-1}$. This large value, which in turn leads to short lifetimes, can be explained by the fact that the entropy of the transition state tends to be larger than the entropy of the initial hopfion state, opposite to what has been found for the collapse of skyrmions~\cite{bessarab:2018}.
Furthermore, they found that different values of the exchange parameters $J_s$ chosen for the spin lattice model~\eqref{eq:heisenberg_hamiltonian} can give slight variations in the size and energy barrier of the hopfion even when the values of the corresponding micromagnetic model parameters $\mathcal{A}, \mathcal{B}$ and $\mathcal{C}$ of Eq.~\eqref{eq:energy_continuous} are the same. The calculations were, however, carried out for hopfions close to the limit of stability, where the energy barrier for collapse is very small, and this may make such variations more pronounced.

\section{Conclusions}
Saddle points for Hopf charge $Q_\mathrm{H} = 1$ toroidal hopfions in cubic bulk magnets with competing exchange interactions have been computed in a systematic fashion on the basis of a classical Heisenberg model evaluated numerically using atomistic spin simulations. The energy barrier associated to the saddle point is an important quantity for the stability of hopfions against thermal excitations.
The reference for our investigation is the RKBDMB-model~\cite{rybakov_magnetic_2019}, which introduces a micromagnetic energy functional for continuous magnetization fields that predicts hopfions in certain magnets with competing exchange interactions. To this functional we have introduced a system of reduced units, which helps separate the influence of the size, symmetry, and energy scale of the hopfions and to cover a wide range of magnetic interaction parameters in the numerical analysis of the hopfion properties.

Based on this reduced unit system, we chose atomistic exchange interaction parameters in an effective four-shell Heisenberg model and calculated the hopfion size, energy and the energy barriers for the collapse of hopfions to either a metastable globule state or the ferromagnetic ground state.
The calculations are carried out using the atomistic spin simulation framework \texttt{Spirit}~\cite{muller_spirit_2019} with an extension of its capabilities by implementing an adapted dimer method for saddle points searches. This method can achieve almost arbitrary resolution around the saddle points while using only two copies of the spin system. By combining the geodesic nudged elastic band (GNEB) method with the dimer method, many saddle points could be computed to higher precision and with less computational effort than would have been possible with the GNEB method alone.
Consistent with the RKBDMB-model~\cite{rybakov_magnetic_2019}, the determined competing Heisenberg exchange interaction parameters led to stable hopfions when the energy minimization was initialized with the proper ansatz function.
In the parameter range investigated, $r_0/a \leq 5$, the hopfion energy is only weakly dependent on the characteristic length scale $r_0$ given in units of the lattice constant $a$, but the energy barriers scale nearly linearly with $r_0$. Both depend on the symmetry parameter $\gamma$. All saddle points found correspond to the formation of pairs of Bloch points in the core of the hopfion and we find that the main property determining the energy barrier for collapse is the size of the hopfion compared to the lattice constant of the magnetic crystal.  
For larger hopfion than those investigated here, $r_0/a > 5$, other collapse mechanisms could come into play and dominate the decay of the hopfion to the ferromagnetic state thus breaking the near linear relationship between energy barrier and $r_0/a$. It seems plausible that the favored collapse mechanism in these cases would mediate the decay via the formation of multiple Bloch points along the circumference of the toroid ring.

By studying toroidal hopfions with Hopf index $Q_\mathrm{H}=1$ as an example, we show how properties of static atomic-scale hopfions in magnets with frustrated Heisenberg exchange interactions can be calculated, including the energy barrier for decay to the ferromagnetic ground state.
Considering the mostly uncharted territory of hopfion stability, and the embedding of this example into its larger context, this work may stimulate the investigation of hopfion stability of more complex and interesting hopfions with different Hopf charge. 

\begin{acknowledgments}
The authors thank V. M. Uzdin, I. S. Lobanov, N. Kiselev,  P. F. Bessarab, G. Kwiatkowski, H. Schrautzer, R. Goswami, M. H. A. Badarneh, M. Hoffmann and G.P. M\"uller for fruitful discussions.
S.B.\  acknowledges funding from Deutsche Forschungsgemeinschaft (DFG) through SPP 2137 ``Skyrmionics'' (grant no.\  BL 444/16-2) and through the Collaborative Research Center SFB 1238 (Project C01) as well as the funding under Helmholtz-RSF Joint Research Group ``TOPOMANN'' and the  European Research Council (ERC) under the European Union’s Horizon 2020 research and innovation programme under grant agreement 856538 (project "3D MAGIC").
H.J.\  acknowledges funding from the Icelandic Research Fund (grant no.\ 185405-053).
\end{acknowledgments}

\appendix
\section{Heisenberg Exchange Parameters}
\label{App:Exchange_Parameters}
In a simple cubic lattice with lattice constant $a$, the relationship between $\mathcal{A}, \mathcal{B}$ and $\mathcal{C}$ of Eq.~\eqref{eq:energy_continuous} and $J_s$ of Eq.~\eqref{eq:heisenberg_hamiltonian} is given by Ref.~\cite{rybakov_magnetic_2019} as
\begin{equation}
    \begin{pmatrix}
        \mathcal{A}\\
        \mathcal{B}\\
        \mathcal{C}
    \end{pmatrix}
    =
    \begin{pmatrix}
        \frac{1}{2a}  & \frac{2}{a} & \frac{2}{a} & \frac{2}{a}\\[4pt]
        -\frac{a}{96} & -\frac{a}{24} & -\frac{a}{24} & -\frac{a}{6}\\[4pt]
        -\frac{a}{48} & -\frac{a}{3} & -\frac{7a}{12} & -\frac{a}{3}
    \end{pmatrix}
    \begin{pmatrix}
        J_1\\
        J_2\\
        J_3\\
        J_4
    \end{pmatrix}.
    \label{eq:abc_from_j}
\end{equation}

Fixing the value of $J_1$ enables us to invert the relationship resulting in
\begin{equation}
        \begin{pmatrix}
            J_2\\
            J_3\\
            J_4
        \end{pmatrix}
        =
        \begin{pmatrix}
            \frac{2}{a} & \frac{2}{a} & \frac{2}{a}\\[4pt]
            -\frac{a}{24} & -\frac{a}{24} & -\frac{a}{6}\\[4pt]
            -\frac{a}{3} & -\frac{7a}{12} & -\frac{a}{3}
        \end{pmatrix}^{-1}
       \begin{pmatrix}
            \mathcal{A} - \frac{J_1}{2a}\\[4pt]
            \mathcal{B} + \frac{J_1a}{96}\\[4pt]
            \mathcal{C} + \frac{J_1a}{48}{C}
        \end{pmatrix}.
        \label{eq:j_from_abc}
\end{equation}

Finding the values for the micromagnetic constants $\mathcal{A}, \mathcal{B}$ and $\mathcal{C}$, given a certain set of reduced units $r_0, \gamma$ and $E_0$, is straightforward as one just inverts Eqs. \eqref{eq:reduced_r}, \eqref{eq:reduced_e} and \eqref{eq:reduced_gamma}:

\begin{equation}
    \begin{pmatrix}
        \mathcal{A}\\
        \mathcal{B}\\
        \mathcal{C}
    \end{pmatrix}
    = E_0 
    \begin{pmatrix}
        \frac{1}{r_0}\\
        r_0 (1-\gamma)\\
        r_0 \gamma
    \end{pmatrix}
    \label{eq:abc_from_reduced}
\end{equation}

Finally, we find the parameters $J_s$, for a given pair of $\gamma$ and $r_0$, by setting $J_1 = 40\,\mathrm{meV}$, $a = 1\,\textup{\AA}$, $E_0 = 1\,\mathrm{meV}$ and using Eqs.~\eqref{eq:j_from_abc} and \eqref{eq:abc_from_reduced}.

\section{Hue-Saturation Color Code for Spin Orientations}
\label{App:Colorcode}
\begin{figure}[!t]
    \centering
    \includegraphics[width=0.7\linewidth]{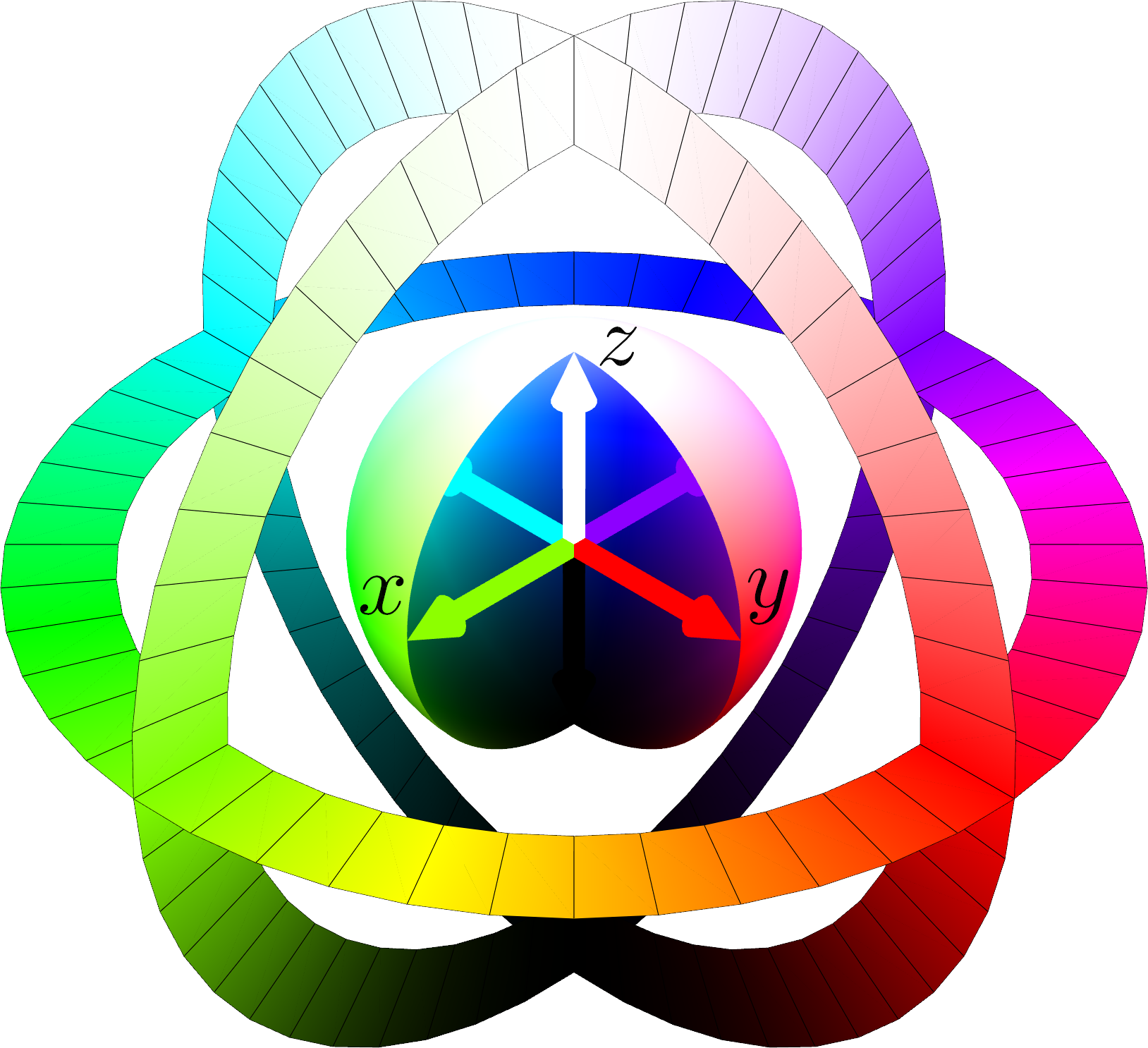}
    \caption{Color code that maps the direction of the spin unit vector to hue and saturation. The azimuthal angle of the spin direction corresponds to the hue, while the polar angle determines the saturation. The spherical shape at the center of the figure is colored according to the radial vector of points on the surface. The surrounding rings show colors according to spins oriented in the $xy$-, $yz$- or $xz$- plane respectively.}
    \label{fig:color_sphere_large}
\end{figure}

Fig.~\ref{fig:color_sphere_large} displays the color code used throughout the article to indicate the direction of the spin unit vectors $\{\vec{n}_i\}$. The frequently shown $n_z = 0$ isosurfaces only make use of the colors of the $xy$-plane ring.

\section{Ansatz for Initial Hopfion Structures}
\label{App:Ansatzfunction}
For the initial guess of the hopfion structures, \textit{i.e.} before energy minimization, the preceding stabilized hopfion with the next largest value of the length scale $r_0$ and the same symmetry parameter $\gamma$ was used.
For the largest value $r_0 = 5a$, for which no preceding minimised hopfion was available, a continuous ansatz field~\cite{Rybakov_Ansatz} for an isotropic toroidal hopfion in a ferromagnetic background aligned to the $n_z$ direction has been used:
\begin{align}
    \vec{n}(\vec{r})=\vec{n}(r,\phi,\theta) =
        \begin{pmatrix}
            \sin t \cos f\\
            \sin t \sin f\\
            \cos t
        \end{pmatrix}, 
\end{align}
where $r$, $\phi$ and $\theta$ are the radius, azimuthal and polar angle of the position vector respectively and
\begin{align}
    t(r,\theta) &= \arccos{\left(-2 \sin^{2}{\theta} \sin^{2}{\xi} + 1 \right)},\\
    f(r,\phi,\theta) &= \phi - \arctan{\left(\frac{1}{\cos{\theta } \tan{\xi}} \right)},\\
    \xi(r) &= \frac{c_1 }{\sqrt{\frac{R^{2}}{r^{2}} + c_1^2}}\,\pi,
\end{align}
with $ c_1=0.4867$ and $R=5a$. 
This ansatz field~\cite{Rybakov_Ansatz} was implemented in the \texttt{Spirit}~\cite{muller_spirit_2019} framework prior to the present work.

\section{Globule Configurations}
\label{App:Globule_configurations}
\begin{figure}[!t]
    \centering
    \includegraphics[width=\linewidth]{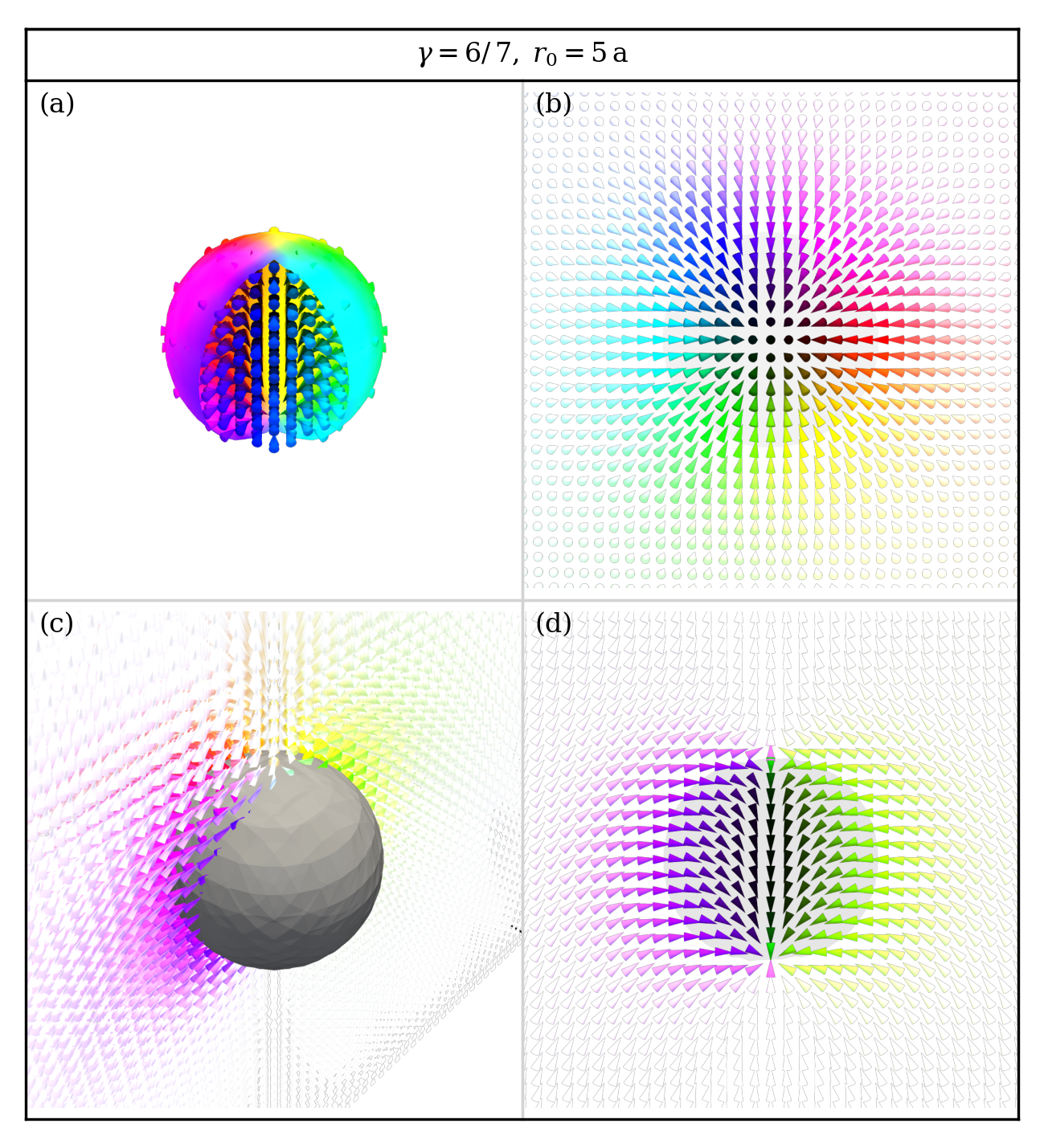}
    \caption{Several depictions of the first globule state encountered in the collapse of the isotropic hopfion with $\gamma = 6/7$ and $r_0 = 5\,a$. (a) The $n_z=0$ isosurface and arrows for any spin with $n_z<0$ (b) Cut through the center of the system orthogonal to the orientation of the core spins.
    (c) The uncolored $n_z=0$ isosurface and a cut through the spin system.
    (d) Cut through the center of the system parallel to the orientation of the core spins.}
    \label{fig:globule_large}
\end{figure}

An enlarged depiction of the first globule state encountered along the MEP of Fig.~\ref{fig:meps}, is shown in Fig.~\ref{fig:globule_large}. It corresponds to the state labelled 'D' in Fig.~\ref{fig:meps} and occurs after the Bloch points have emerged and the outer structure of the hopfion has contracted. The $n_z = 0$ isosurface has a spherical structure and consists of two Bloch points at the poles which are connected by a string of spins oriented anti-parallel to the background spin orientation.

\section{Determination of Hopfion Size}
To compute the hopfion radius $R_\mathrm{H}$, used in Fig.~\ref{fig:energy_barriers}, the following scheme has been applied:
First, the center point of the hopfion $\vec{C}$ was found by taking the average of all lattice positions where the angle between the spin direction $\vec{n}_i$ and the direction of the ferromagnetic background exceeded $\pi/4$,
\begin{equation}
    \vec{C} = \big\langle \vec{r}_i \big\rangle \text{ where } \vec{r}_i \in \mathcal{R}_{\pi/4}
\end{equation}
with 
\begin{equation}
    \mathcal{R}_{\pi/4} = \big\{ \vec{r}_i \text{ where } \angle(\vec{n}_i,\nvec{e}_z) > \pi/4 \big\}.
\end{equation}
Then, the direction of the hopfion normal $\vec{N}$ was determined by numerically minimizing the objective function
\begin{equation}
    K(\phi,\theta) = \sum_{\vec{r}_i \in \mathcal{R}_{\pi/4}} \left[\vec{N}(\phi,\theta)\cdot (\vec{r}_i-\vec{C})\right]^2,
\end{equation}
with
\begin{equation}
\vec{N}(\phi,\theta) = 
    \begin{pmatrix}
        \cos \phi \sin \theta\\
        \sin \phi \sin \theta\\ 
        \cos \theta
    \end{pmatrix}.
\end{equation}

Finally, we evaluate $R_\mathrm{H}$ by taking the average over the position vectors in the $n_z=0$ isosurface, projected onto the equatorial plane:
\begin{equation}
    R_\mathrm{H} = \big\langle \left|\vec{r}_i - (\vec{r_i}\cdot\vec{N})\vec{N} \right| \big\rangle
    \text{ with } (\vec{n}_i)_z \in [-0.1, 0.1.]
\end{equation}

\FloatBarrier
\bibliography{references}
\end{document}